\documentclass[aip,apl,reprint]{revtex4-1}
\usepackage{amsfonts}
\usepackage{amsmath}
\usepackage{amssymb}
\usepackage{graphicx}

\usepackage[bookmarks=true,bookmarksopen,bookmarksnumbered,
                colorlinks,
                linkcolor=blue,
                citecolor=blue,breaklinks=true]{hyperref}

\begin{document}

\title{Inductive-detection electron-spin resonance spectroscopy with $\mathbf{65}\,$spins$/\sqrt{\text{Hz}}$ sensitivity} 

\author{S.~Probst}
\email[]{sebastian.probst@cea.fr}

\affiliation{Quantronics group, SPEC, CEA, CNRS, Universit\'e Paris-Saclay, CEA Saclay 91191 Gif-sur-Yvette Cedex, France}

\author{A.~Bienfait}
\affiliation{Quantronics group, SPEC, CEA, CNRS, Universit\'e Paris-Saclay, CEA Saclay 91191 Gif-sur-Yvette Cedex, France}
\affiliation{Institute for Molecular Engineering, University of Chicago, Chicago, Illinois 60637, USA}

\author{P.~Campagne-Ibarcq}
\affiliation{Quantronics group, SPEC, CEA, CNRS, Universit\'e Paris-Saclay, CEA Saclay 91191 Gif-sur-Yvette Cedex, France}
\affiliation{Departments of Applied Physics and Physics, Yale University, New Haven, CT 06520, USA}

\author{J.~J.~Pla}
\affiliation{School of Electrical Engineering and Telecommunications, University of New  South Wales, Anzac Parade, Sydney, NSW 2052, Australia}

\author{B.~Albanese}
\affiliation{Quantronics group, SPEC, CEA, CNRS, Universit\'e Paris-Saclay, CEA Saclay 91191 Gif-sur-Yvette Cedex, France}

\author{J.~F.~Da~Silva~Barbosa}
\affiliation{Quantronics group, SPEC, CEA, CNRS, Universit\'e Paris-Saclay, CEA Saclay 91191 Gif-sur-Yvette Cedex, France}

\author{T.~Schenkel}
\affiliation{Accelerator Technology and Applied Physics Division, Lawrence Berkeley National Laboratory, Berkeley, California 94720, USA}

\author{D.~Vion}
\affiliation{Quantronics group, SPEC, CEA, CNRS, Universit\'e Paris-Saclay, CEA Saclay 91191 Gif-sur-Yvette Cedex, France}

\author{D.~Esteve}
\affiliation{Quantronics group, SPEC, CEA, CNRS, Universit\'e Paris-Saclay, CEA Saclay 91191 Gif-sur-Yvette Cedex, France}

\author{K.~M{\o}lmer}
\affiliation{Department of Physics and Astronomy, Aarhus University, Ny Munkegade 120, DK-8000 Aarhus C, Denmark}

\author{J.~J.~L.~Morton}
\affiliation{London Centre for Nanotechnology, University College London, London WC1H 0AH, United Kingdom}

\author{R.~Heeres}
\affiliation{Quantronics group, SPEC, CEA, CNRS, Universit\'e Paris-Saclay, CEA Saclay 91191 Gif-sur-Yvette Cedex, France}

\author{P.~Bertet}
\affiliation{Quantronics group, SPEC, CEA, CNRS, Universit\'e Paris-Saclay, CEA Saclay 91191 Gif-sur-Yvette Cedex, France}

\date{\today}

\begin{abstract}
We report electron spin resonance spectroscopy measurements performed at millikelvin temperatures in a custom-built spectrometer comprising a superconducting micro-resonator at $7$\,GHz and a Josephson parametric amplifier. Owing to the small ${\sim}10^{-12}\lambda^3$ magnetic resonator mode volume and to the low noise of the parametric amplifier, the spectrometer sensitivity reaches $260\pm40$\,spins$/$echo and $65\pm10\,\mathrm{spins}/\sqrt{\text{Hz}}$, respectively. 
\end{abstract}


\maketitle 

\newcommand{\pb}{\textcolor{red}}
\newcommand{\seb}{\textcolor{green}}

Electron spin resonance (ESR) is a well-established spectroscopic method to analyze paramagnetic species, utilized in materials science, chemistry and molecular biology to characterize reaction products and complex molecules \cite{SchweigerEPR(2001)}. In a conventional ESR spectrometer based on the so-called inductive detection method, the paramagnetic spins precess in an external magnetic field $B_0$ and radiate weak microwave signals into a resonant cavity, whose emissions are amplified and measured.\\

Despite its widespread use, ESR has limited sensitivity, and large amounts of spins are necessary to accumulate sufficient signal. Most conventional ESR spectrometers operate at room temperature and employ three-dimensional cavities. At X-band\footnote{X-band frequency range: 8 to 12\,GHz.}, they require on the order of ${\sim}10^{13}$ spins to obtain sufficient signal in a single echo \cite{SchweigerEPR(2001)}.
Enhancing this sensitivity to smaller spin ensembles and eventually the single-spin limit is highly desirable and is a major research subject.
This has been achieved by employing alternative detection schemes including optically detected magnetic resonance (ODMR) \cite{Wrachtrup1993,Gruber1997}, scanning probe based techniques \cite{Baumann2015,Manassen1989,Rugar1992,Rugar2004,Grinolds2014}, SQUIDs \cite{Chamberlin1979} and electrically detected magnetic resonance \cite{Hoehne2012,Morello2010}. For instance, ODMR achieves single spin sensitivity through optical readout of the spin state. However, this requires the presence of suitable optical transitions in the energy spectrum of the system of interest, which makes it less versatile.

In recent years, there has been a parallel effort to enhance the sensitivity of inductive ESR detection \cite{Narkowicz2008,Shtirberg2011,Kubo2012,Malissa2013,Sigillito.APL2014,bienfait2015reaching,Bienfait2016squeezing,Eichler2017}. This development has been triggered by the progress made in the field of circuit quantum electrodynamics (cQED)\cite{Devoret2013}, where high fidelity detection of weak microwave signals is essential for the measurement and manipulation of superconducting quantum circuits. In particular, it has been theoretically predicted~\cite{Haikkasinglespin2017} that single-spin sensitivity should be reachable by combining high quality factor superconducting micro-resonators and Josephson Parametric Amplifiers (JPAs)\cite{Zhou2014}, which are sensitive microwave amplifiers adding as little noise as allowed by quantum mechanics to the incoming spin signal. Based on this principle, ESR spectroscopy measurements\cite{bienfait2015reaching} demonstrated a sensitivity of $1700$\,spins$/\sqrt{\text{Hz}}$. In this work, we build on these efforts and show that, by optimizing the superconducting resonator design, the sensitivity can be enhanced to the level of $65\,$spins$/\sqrt{\text{Hz}}$.

\begin{figure}
	\centering
	\includegraphics{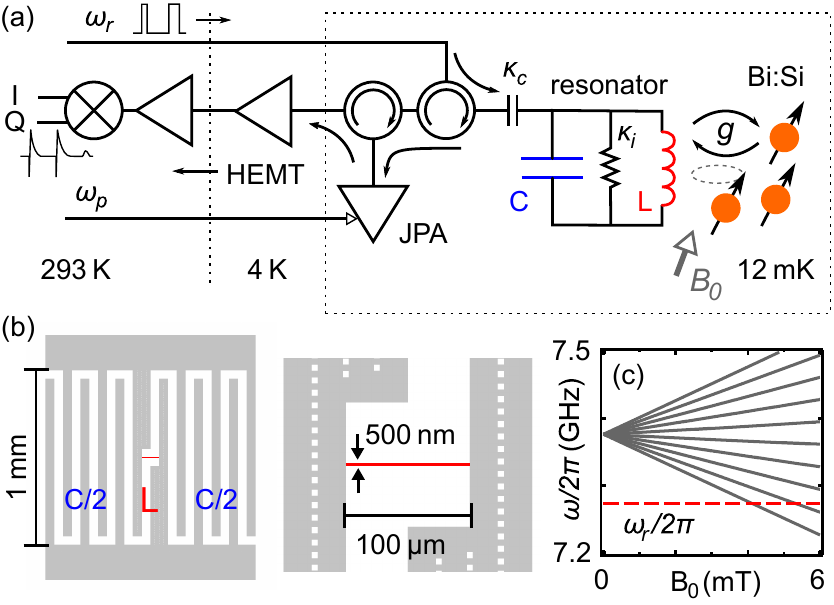}%
	\caption{\label{fig1} (a) Schematic of the experiment: Bi:Si spins, biased by a dc magnetic field $B_0$, are coupled to a LC resonator of frequency $\omega_r$. Microwave control pulses at $\omega_r$ are sent onto the resonator input. The reflected signal, as well as the signals emitted by the spins, are first amplified by a JPA operated in degenerate mode followed by further amplification and homodyne demodulation to obtain the signal quadratures $I(t)$ and $Q(t)$. (b) Design of the planar lumped element LC resonator. (c) ESR-allowed transitions of the Bi donor spins vs.~$B_0$. Dashed line indicates the resonator frequency.}%
\end{figure}

Figure \ref{fig1}(a) shows a schematic design of the spectrometer consisting of a superconducting LC resonant circuit capacitively coupled to the measurement line with rate $\kappa_c$ and internal losses $\kappa_i$. The resonator is slightly over-coupled ($\kappa_c \gtrsim \kappa_i$) and probed in reflection at its resonance frequency $\omega_r$. This micro-resonator is inductively coupled to the spin ensemble and cooled to $12$\,mK in a dilution refrigerator. The signal leaking out of the resonator, which contains in particular the spin signal, is first amplified by a JPA operating in the degenerate mode\cite{Caves1982,Yamamoto2008}, followed by a High-Electron-Mobility Transistor (HEMT) amplifier at 4\,K and further amplifiers at room-temperature. The two signal quadratures $I(t)$ and $Q(t)$ are obtained by homodyne demodulation at $\omega_r$. More details on the setup can be found in Ref.~\onlinecite{bienfait2015reaching}.

Compared to Ref.~\onlinecite{bienfait2015reaching}, the micro-resonator was re-designed with the goal of enhancing the spin-resonator coupling constant $g=\gamma_e\left< 0 \right| S_x \left| 1 \right> \delta B_1$, where $\left< 0 \right| S_x \left| 1 \right>\approx0.5$ for the transition used in the following. Here, $\gamma_e/2\pi=28\,$GHz$/$T denotes the gyromagnetic ratio of the electron, $\left| 0 \right>$ and $\left| 1 \right>$ the ground and excited state of the spin, $\mathbf{S}$ the electron spin operator and $\delta B_1$ the magnetic field vacuum fluctuations. Reducing the inductor size to a narrow wire decreases the magnetic mode volume\cite{HarocheBook} and therefore enhances $\delta B_1$.
In the new design, shown in Fig.~\ref{fig1}b, most of the resonator consists of an interdigitated capacitor, shunted by a $l=100$\,$\mu$m long, $w=500\,$nm wide, and $t=100\,$nm-thick wire inductance. It is patterned out of an aluminum thin-film by electron-beam lithography followed by lift-off, on top of an isotopically enriched $^{28}$Si sample containing bismuth donors implanted at a depth of $z \approx 100$\,nm.
Based on electromagnetic simulations, an impedance of $32\,\Omega$ and a magnetic mode volume of ${\sim}10^{-12}\lambda^3$ ($0.2\,$pico-liters) are estimated, resulting in a spin-resonator coupling of $g/2\pi\approx 4.3\cdot 10^2$\,Hz.
The resonator properties are characterized at $12$\,mK by microwave reflection measurements\cite{PozarMicrowave,Probst2015}, yielding $\omega_r/2\pi = 7.274$\,GHz, $\kappa_c = 3.4\cdot 10^5\,$rad\,s$^{-1}$, $\kappa_i = 2.5\cdot 10^5\,$rad\,s$^{-1}$ and a total loss rate of $\kappa_l = \kappa_i + \kappa_c = 5.9\pm0.1\cdot 10^5\,$rad\,s$^{-1}$, measured at a power corresponding to a single photon on average in the resonator\cite{Martinis2008}. \\

At low temperatures, bismuth donors in the silicon sample trap an additional valence electron to the surrounding host silicon atoms, which can be probed through electron spin resonance.~\cite{Feher.PhysRev.114.1219(1959),Morley.NatureMat.9.725(2010)}. The electron spin $S=1/2$ experiences a strong hyperfine interaction ($A/2\pi=1.45\,$GHz) with the $^{209}$Bi nuclear spin $I=9/2$ giving rise to a zero field splitting of $7.38\,$GHz. The full Hamitonian is given by
$H/\hbar = \gamma_e \,\mathbf{S} \cdot \mathbf{B} - \gamma_n \mathbf{I} \cdot \mathbf{B} + A \,\mathbf{S}\cdot \mathbf{I}$\,,
where $\gamma_n/2\pi=7\,$MHz$/$T denotes the gyromagnetic ratio of the nucleus. Note that the Bi spin system is also interesting in the context of quantum information processing because it features clock transitions where the coherence time can reach $2.7\,$s \cite{Wolfowicz2013}. In addition, the large zero field splitting makes this system well suited for integration with superconducting circuits. Figure \ref{fig1}(c) shows the low field spectrum of the ESR-allowed transitions close to the resonator frequency. The dashed line marks the spectrometer resonator frequency at $\omega_r/2\pi=7.274\,$GHz.
\\

\begin{figure}
	\centering
	\includegraphics{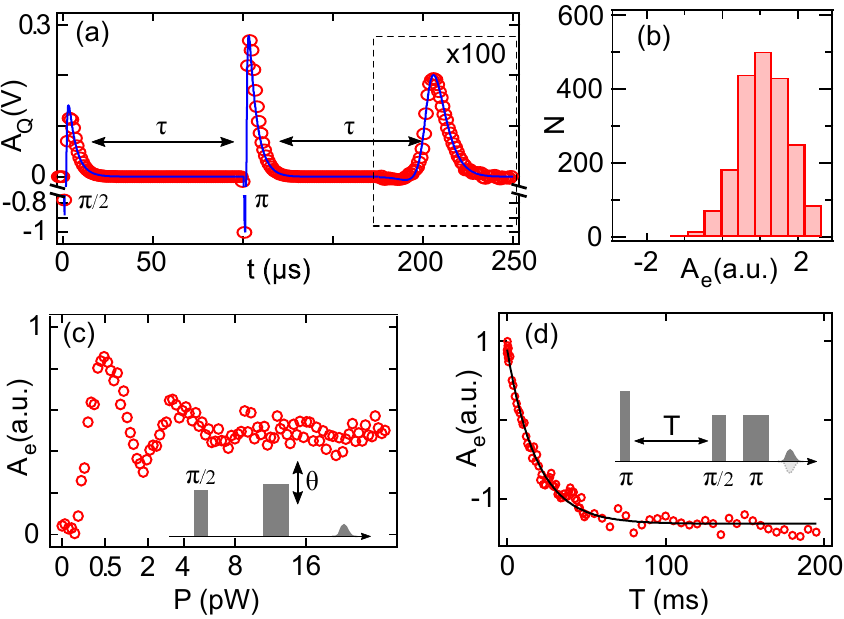}%
	\caption{\label{fig2} (a) Measured (red circles) and simulated (blue line) quadrature signal showing the $\pi$ and $\pi/2$ pulses as well as the echo. (b) Histogram of $A_e$. These data are obtained by subtracting two consecutive experimental traces with opposite $\pi$/2 pulse phases (phase cycling\cite{bienfait2015reaching}), so that the single-echo SNR is obtained from the histogram width multiplied by $\sqrt{2}$. (c) Rabi oscillations of $A_e$, recorded by varying the power of the second pulse of the spin echo sequence. (d) Spin relaxation time measurement. $A_e$ measured as a function of the delay $T$ between an initial $1\,\mu$s-long $\pi$ pulse and a subsequent spin-echo sequence (red open circles). An exponential fit (black solid line) yields $T_1=18.6\,$ms.}%
\end{figure}

For the sensitivity of the spectrometer, two quantities are relevant: the minimum number of spins $N_{min}$ necessary to produce a single echo with a signal-to-noise ratio (SNR) of 1, as well as the number of spins that can be measured with unit SNR within $1$ second of integration time $N_{min} / \sqrt{N_{seq}}$ where $N_{seq}$ is the number of experimental sequences per second. This timescale is determined by the spin energy relaxation time $T_1$, and we typically wait $T_\text{rep}\gtrsim 3 T_1$ between measurements.
In our experiment, the lowest transition of the Bi ensemble is tuned into resonance with the cavity by applying $B_0=3.74\,$mT parallel to the central inductor. In order to address all spins within the cavity bandwidth, we choose the duration $t_p$ of our square pulses $0.5\,\mu$s for the $\pi/2$ and $1\,\mu$s for the $\pi$ pulse such that $t_p\,\kappa_l \lesssim 1$. The $\pi$ pulse amplitude was determined by recording Rabi oscillations on the echo signal, see Fig.~\ref{fig2}(c). Figure \ref{fig2}(a) shows a full echo sequence (red circles). The reflected control pulses show a rapid rise followed by a slower decay due to the resonator ringdown, leading to an asymmetric echo shape.

In order to simulate the data, knowledge of $g$ is necessary~\cite{bienfait2015reaching}. It is experimentally obtained from spin relaxation data, as explained in the next paragraph, leaving no other adjustable parameter than the number of spins excited by the first $\pi/2$ pulse. The quantitative agreement, see blue line in Fig.~\ref{fig2}(a), allows us to state that $N_e=234\pm35\,$spins are contributing to the echo. $N_e$ is defined through the polarization created by the first $\pi/2$ pulse. For details on the simulation we refer to Ref.~\onlinecite{bienfait2015reaching}. The ESR signal is given by the echo area $A_e$ and in order to extract the SNR, a series of echo traces was recorded. Each echo trace is then integrated, weighted by its expected mode shape, which constitutes a matched filter maximizing the SNR\cite{bienfait2015reaching}. From the resulting histogram, shown in Fig.~\ref{fig2}(b), we deduce a SNR of $0.9$ per single trace, yielding a single shot sensitivity of $N_\text{min}=260\pm40$ spins per echo.
This result is consistent with an estimate of $N^\text{(th)}_\text{min}= \frac{\kappa_l}{2gp} \sqrt{\frac{n w}{\kappa_c}}\approx 10^2$\,spins using the theory developed in Ref.~\onlinecite{bienfait2015reaching}.
Here, $n=0.5$ is the number of noise photons, $p=1-\exp(-3T_1/T_1)$ the polarization and $w\approx \kappa_l$ the effective inhomogeneous spin linewidth.
Since the experiment was repeated at a rate of $16\,$Hz, this single echo sequence translates into an absolute sensitivity of $65\pm10\,$spins$/\sqrt{\text{Hz}}$. This figure may be increased further by irradiating the resonator with squeezed vacuum, as demonstrated in Ref.~\onlinecite{Bienfait2017squeezing}.\\

Figure \ref{fig2}(d) shows the longitudinal decay of the spin ensemble.
It was obtained with an inversion recovery pulse sequence: first, a $1\,\mu$s-long $\pi$ pulse inverts the spin ensemble followed by a spin echo detection sequence with $5\,\mu$s and $10\,\mu$s-long pulses after a variable time $T$. The exponential fit yields $T_1 = 18.6\pm0.5\,$ms. As shown in Ref.~\onlinecite{bienfait2016Purcell}, the energy relaxation of donors in silicon coupled to small-mode-volume and high-quality-factor resonators is dominated by spontaneous emission of microwave photons into the environment, at a rate $T_1^{-1} = 4 g^2 / \kappa_l$. This allows us to experimentally determine that $g/2\pi = 450\pm11$\,Hz, which is close to the value estimated from design.

With the current sensitivity of $65\,$spins$/\sqrt{\text{Hz}}$, more than 1 hour of integration time would be needed to measure a single spin with unit SNR.
Since the integration time needed to accumulate a signal with a given SNR scales proportional to $g^{-4}$ as explained in Ref.~\onlinecite{Haikkasinglespin2017}, increasing the coupling constant by one order of magnitude would be sufficient to obtain single-spin sensitivity in less than a second integration time.
This can be achieved by bringing the spins closer to the inductor of the resonator using an even thinner and narrower inductor to concentrate $\delta B_1$, and by reducing the impedance of the resonator further\cite{Eichler2017}. 
\\

\begin{figure}
	\centering
	\includegraphics{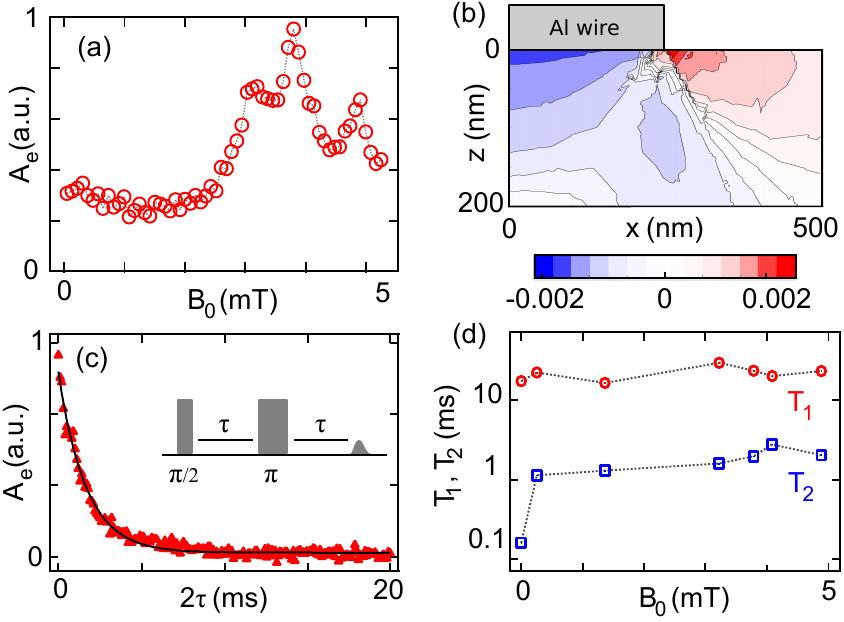}%
	\caption{\label{fig3} (a) Echo-detected field sweep. $A_e$ (open circles) is shown as a function of $B_0$ (parallel to the wire). (b) COMSOL\textsuperscript \textregistered simulation of the $\epsilon_{100}$ component of the strain field in the silicon around the wire. (c) Spin coherence time measurement at $B_0=3.74\,$mT. $A_e$ plotted as a function of the delay $2\tau$ between $\pi/2$ pulse and echo (red triangles). An exponential fit (black solid line) yields $T_2=1.65\pm0.03\,$ms. (d) $T_1$ and $T_2$ as a function of $B_0$. Error bars are within the marker size.}%
\end{figure}

Figure \ref{fig3}(a) displays a Hahn-echo field sweep, i.e. $A_e$ as a function of $B_0$ applied parallel to the inductor. The curve shows a large inhomogeneous broadening with Bi spins detected even at $B_0 = 0\,$mT, which are thus shifted by approximately $100\,$MHz from the nominal zero-field value, see Fig.~\ref{fig1}(c). We attribute this broadening to strain exerted by the aluminum resonator onto the Si substrate resulting from a difference in their coefficients of thermal expansion\cite{Pla2016,Zimmermann2015,bienfait2015reaching}. Figure~\ref{fig3}(b) displays a COMSOL\textsuperscript \textregistered simulation of the $\epsilon_{100}$ component of the strain tensor. The impact of strain on the Bi spectrum is subject of active experimental and theoretical research\cite{Pla2016,Mansir2017}.
We have investigated the dependence of the spin coherence and relaxation times on $B_0$, as shown in Fig.~\ref{fig3}(d). A typical coherence time measurement, recorded at $B_0 = 3.74$\,mT by measuring $A_e$ as a function of $2\tau$, is shown in Fig.~\ref{fig3}(c). The data are well fitted by an exponential decay with $T_2=1.65\pm0.03\,$ms. While $T_1$ shows nearly no dependence on $B_0$, $T_2$ decreases weakly towards lower magnetic fields and drops abruptly at zero field. This behavior might be due to fast dynamics within the bismuth donor Zeeman sub-levels induced at low fields by a residual concentration of $^{29}\mathrm{Si}$ nuclear spins, although more work is needed to draw a definite conclusion.\\

\begin{figure}
	\centering
	\includegraphics{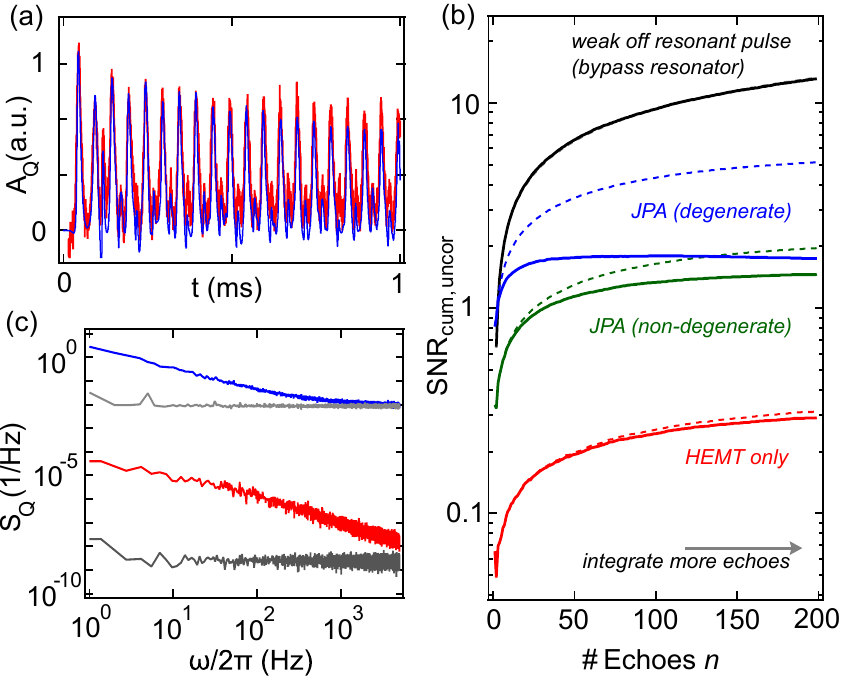}%
	\caption{\label{fig4} (a) Averaged quadrature signal (red solid line) and simulation (blue solid line) showing the echoes recorded during the first millisecond of the CPMG sequence. (b) SNR vs.~number of averaged CPMG echoes employing just the HEMT amplifier, the JPA in non-degenerate mode, the JPA in degenerate mode and a control experiment, see text for details. Solid lines show the data, dashed lines the expected gain in SNR assuming uncorrelated noise. (c) Normalized quadrature noise power spectrum $S_Q(\omega)$ of the resonator at high (red) and low (blue) power corresponding to an average population of $10^6$ and $3$ photons in the cavity, respectively. Both bright and dark gray traces show the corresponding off-resonant noise traces for comparison.}%
\end{figure}

The sensitivity of the current spectrometer can be further enhanced by using multiple refocusing pulses to generate several echoes per sequence. Here, we employ the Carr-Purcell-Meiboom-Gill (CPMG) sequence \cite{SchweigerEPR(2001),CPMG}, which consists of a $\pi/2$ pulse applied along the x-axis followed by $n$ $\pi$ pulses along the y-axis of the Bloch sphere. 
Assuming uncorrelated Gaussian noise, the increase of SNR is given by the CPMG echo decay curve $\text{SNR}(n)/\text{SNR}(1)=\frac{1}{\sqrt{n}}\sum_{i=1}^n{A_e(t_i)}$, where the index $i$ labels the echoes from $1$ to $n$ along the sequence. The individual echoes during the first millisecond are presented in Fig.~\ref{fig4}(a). The refocusing pulses are not visible in this plot because they are canceled by phase cycling. The blue line, computed by the simulation presented in Fig.~\ref{fig2}(a) and using the same system parameters, is in good agreement with the data.

In order to quantify the gain in SNR, we record up to $4\cdot 10^4$ single CPMG traces containing 200 echoes each. The data are then analyzed in two ways presented in Fig.~\ref{fig4}(b) by dashed and solid lines, respectively: First, each echo in each sequence is integrated individually and its mean $\bar{x}_i$ and standard deviation $\Delta x_i$ are calculated in order to determine the $\text{SNR}_i=\bar{x}_i/\Delta x_i$ of the $i$-th echo. Provided that the noise is uncorrelated, the cumulative SNR sum over $n$ echoes is given by $\text{SNR}_\text{uncor}=\frac{1}{\sqrt{n}}\sum_{i=1}^n \text{SNR}_i$. Second, we determine the actual cumulative $\text{SNR}_\text{cum}=\bar{x}_\text{cum}/\Delta x_\text{cum}$ by summing up all echoes in each trace up to the $n$-th echo and subsequently calculate the mean  and standard deviation. Figure~\ref{fig4}(b) shows the result for the spectrometer operating just with a HEMT amplifier, with the JPA in phase preserving mode and with the JPA in the degenerate mode. Without the JPA, $\text{SNR}_\text{uncor}\approx \text{SNR}_\text{cum}$ yielding a gain in SNR of up to $6$. Employing the JPA, the gain initially follows the expectation for $\text{SNR}_\text{uncor}$ but then saturates. In particular, in the highest sensitivity mode, CPMG only allows for an increase in the SNR by approximately a factor of 2, thus reaching $33\,$spins$/\sqrt{\text{Hz}}$. We interpret the discrepancy between $\text{SNR}_\text{cum}$ and $\text{SNR}_\text{uncor}$ as a sign that correlations exist between the noise on the echoes of a given sequence, or in other words that low-frequency noise is present in our system.\\

To investigate whether this low-frequency noise is caused by the microwave setup (including the JPA), we perform a control experiment by replacing the echoes by weak coherent pulses of similar strength, which are reflected at the resonator input without undergoing any phase shift because they are purposely detuned by ${\sim}25\kappa_l$ from $\omega_r$. Figure~\ref{fig4}(b) shows that $\text{SNR}_\text{uncor}= \text{SNR}_\text{cum}$ for this reference measurement (black dashed and solid lines are superimposed) indicating that the JPA itself is not responsible for the observed low frequency noise.  Instead, we attribute the sensitivity saturation in the echo signal to phase noise of our resonator.
Figure~\ref{fig4}(c) presents the normalized on and off resonance quadrature noise power spectra $S_Q(\omega)$ of the out-of-phase quadrature\cite{Gao2007} for two different powers. The noise originating from the resonator (blue and red line) shows a $S_Q(\omega)\propto 1/\omega$ dependence dominating the background white noise (gray and black line). For the low power measurement (blue line), corresponding to an average population of $3$ photons in the resonator, we obtain a rms frequency noise of $7\,$kHz, which is $7\,$\% of $\kappa_l$. This amount of phase noise is commonly observed in superconducting micro-resonators \cite{Gao2007}. Compared to low power, the high power spectrum (red line), corresponding to an average population of $10^6$ photons, shows significantly less noise and we find that $S_Q(\omega)$ scales with the square-root of the intra-cavity power\cite{Gao2007,Martinis2008}. This suggests that origin of the low frequency excess noise lies in the presence of dielectric and/or paramagnetic defects\cite{deGraaf2017,deGraaf2017b,Paladino2014,Wang2015,Gao2008,Macha2010,Sendelbach2008,Burnett2014,Anton2013}.\\

In conclusion, we have presented spin-echo measurements with a sensitivity of $65\,$spins$/\sqrt{\text{Hz}}$, setting a new state-of-the-art for inductively-detected EPR.
This was obtained by employing a low mode volume planar superconducting resonator in conjunction with a quantum limited detection chain. The energy lifetime of the spins was limited by the Purcell effect to $20\,$ms, allowing for fast repeating measurements. Due to the long coherence time of the spin system under investigation, Bi donors in $^{28}$Si, it was possible to enhance the sensitivity further by a CPMG sequence to $33\,$spins$/\sqrt{\text{Hz}}$. Achieving the maximum theoretical sensitivity with CPMG of $11\,$spins$/\sqrt{\text{Hz}}$ was most likely hindered by the phase noise of the resonator. These experiments present a further step towards single-spin sensitivity, and the sub pico-liter detection volume of our spectrometer makes it an interesting tool for investigating paramagnetic surfaces and, in particular, recently discovered 2D materials\cite{Geim2013,Novoselovc2016}.\\

We acknowledge technical support from P.~S\'enat and P.-F.~Orfila, as well as useful and stimulating discussions within the Quantronics group. We acknowledge support of the European Research Council under the European Community's Seventh Framework Programme (FP7/2007-2013) through grant agreements No.~615767 (CIRQUSS), 279781 (ASCENT), and 630070 (quRAM), and of the ANR project QIPSE as well as the the Villum Foundation.

\bibliography{literature}

\end{document}